\documentclass[fleqn,usenatbib]{mnras}
\usepackage{newtxtext,newtxmath}
\usepackage[T1]{fontenc}
\usepackage{ae,aecompl}
\usepackage{graphicx}	% Including figure files
\usepackage{amsmath}	% Advanced maths commands
\usepackage{amssymb}	% Extra maths symbols
\newcommand{\msun}{\,{\rm M_{\odot}}}
\newcommand{\thobs}{\,{\theta_{\rm obs}}}
\newcommand{\cm}{\,{\rm cm}}
\newcommand{\s}{\,{\rm s}}	
\newcommand{\erg}{\,{\rm erg}}

\newcommand{\A}{{\it A}}
\newcommand{\B}{{\it B}}
\newcommand{\C}{{\it C}}
\newcommand{\D}{{\it D}}

\title[EM signals from the first hours of NS mergers]{Electromagnetic signals from the decay of free neutrons in the first hours of neutron star mergers}

\author[Gottlieb \& Loeb]{
Ore Gottlieb$^{1}$\thanks{oregottlieb@mail.tau.ac.il},
Abraham Loeb$^{2}$
\\
$^{1}${The Raymond and Beverly Sackler School of Physics and Astronomy, Tel Aviv University, Tel Aviv 69978, Israel}\\
$^{2}$Astronomy Department, Harvard University, 60 Garden St., Cambridge, MA 02138, USA
}

\pubyear{2020}

\begin{document}
\label{firstpage}
\pagerange{\pageref{firstpage}--\pageref{lastpage}}
\maketitle

\begin{abstract}

The first hours following a neutron star merger are considered to provide several UV/optical/NIR signals: $ \beta $-decay emission from free neutrons, radioactive decay of shocked heavy elements in the cocoon and cocoon's cooling emission. Here we consider two additional emission sources: $ \beta $-decay of free neutrons in the cocoon and synchrotron by the $ \beta $-decay electrons.
We present 3D RHD simulations of jets that propagate in a multi-layer ejecta from the merger and calculate semi-analytically the resulting light curves.
We find that the free neutrons emission at high latitudes is enhanced by the cocoon by a factor of a few to power a wide ($ \lesssim 60^\circ $) and brief ($ \sim 1 $ hour) UV signal that can reach an absolute magnitude of $\gtrsim$ -15, comparable with the cooling emission.
If the ejected neutron matter mass is $ M_n \gtrsim 10^{-4}\msun $, the synchrotron emission may yield a long ($ \sim 8 $ hours) quasi-isotropic UV/optical signal with an absolute magnitude between -12 and -15, depending on the magnetic field.
Such a high mass of a mildly-relativistic component may partly obscure the cocoon's shocked r-process elements, thereby attenuating its radioactive decay emission.
Future observations on these timescales, including null detections, may place constraints on the ejected neutron matter mass and shed light on the ejecta and jet-cocoon characteristics.

\end{abstract}


\section{Introduction}

The first gravitational wave event from a binary neutron star merger (NSM), GW170817, produced detectable signals throughout the entire electromagnetic spectrum, ranging from the very first seconds to years later.
Following the detections of gravitational waves \citep{Abbott2017a} and $ \gamma $-rays 1.7s later \citep{Goldstein2017,Savchenko2017}, a search for the host galaxy was initiated. But it was only 10.9 hours after the merger that the first optical counterpart was detected and the host galaxy was found \citep{Coulter2017}. Over the course of ten days, additional optical and IR signals were detected and allowed a detailed examination of Kilonova models \citep{Li1998a,Kulkarni2005a,Metzger2010a,Barnes2013a,Kasen2013a,Tanaka2013a}. The observations were found to be in a good agreement with theoretical predictions (see e.g. \citealt{Kasen2017}), suggesting that a significant amount of mass $ M\approx 0.05\msun $ of heavy r-process elements was ejected.
While optical and IR observations of GW170817 shed light on the post-merger evolution on day timescales, the lack of observations in the first 10.9 hours keeps the early evolution of the system in the dark. During this early period, several potential emission sources may yield a detectable signal:
\\
(i) {\bf $ \beta $-decay of free neutrons} \citep{Kulkarni2005a,Metzger2015a}:
As a shock crosses the outer crust of the NS during the coalescence, it may disintegrate heavy nuclei to free neutrons (e.g. \citealt{Ishii2018}). The neutrons may avoid being captured into nuclei if they reach mildly-relativistic velocities so that their interaction with the slower ejected mass of heavy nuclei is minimal \citep{Bauswein2013,Just2015a,Goriely2015,Ishii2018}. Free neutrons can also form in the disc surrounding the newly formed compact object, if the electron fraction is low \citep{Perego2014a}. Simulations of NSMs \citep{Bauswein2013,Just2015a} have found that a relatively large amount of free neutrons is ejected ($ \sim 10^{-4}\msun $), and may power an early bright UV/optical signal \citep{Metzger2015a}, which is possibly polarized \citep{Matsumoto2018a}.
\\
(ii) {\bf Cocoon's cooling emission and heavy r-process $ \beta $-decay} (``cocoon's Kilonova") \citep{Nakar2017,Gottlieb2018a,Kasliwal2017,Piro2018}:
VLBI radio images of GW170817 revealed a superluminal motion of the emitting source between 75 and 230 days after the merger \citep{Mooley2018}. This observation confirmed that NSMs are associated with relativistic jets. The jet, launched a short time after the merger, has to propagate through the expanding ejecta. The jet-ejecta interaction shocks jet and ejecta material to form a hot cocoon around the jet. The cocoon powers a $ \gamma $/X-ray signal on timescales of seconds to minutes when breaking out from the expanding ejecta \citep{Gottlieb2018b}. As the cocoon cools down, it can give rise to a bright UV/optical cooling emission over the course of minutes to hours. After a couple of hours, the mildly-relativistic shocked heavy elements in the cocoon undergo radioactive decay to power the cocoon's Kilonova in the optical/NIR bands \citep{Nakar2017,Gottlieb2018a}.
\\
(iii) {\bf $ \beta $-decay of free neutrons in the cocoon}:
While the origin of the $ \gamma $-ray signal in GW170817 is still under debate, the most likely explanation is a cocoon shock breakout from a mildly-relativistic tail of the ejecta \citep{Gottlieb2018b}. The existence of the tail ejecta has been suggested by several theoretical arguments \citep{Hotokezaka2012,Hotokezaka2018,Kyutoku2012,Bauswein2013,Beloborodov2018,Radice2018}, and is most likely linked to the ejection of free neutrons.
If this component is quasi-isotropic, its interaction with the cocoon is inevitable. Such interaction accelerates free neutrons further and alters their spatial distribution to affect their early emission.
\\
(iv) {\bf Synchrotron emission from the $ \beta $-decay electrons}:
During the decay of the free neutrons, mildly-relativistic electrons are emitted with an energy of $ \sim 0.25 $MeV. If the medium into which the electrons are emitted has some degree of magnetization, the electrons emit synchrotron radiation. The large number of electrons ($ \sim 10^{53} $) emitted during the first hours may power a detectable signal.

A detection of an electromagnetic probe during the first hours after the merger is of a great importance as such a signal may help constraining the composition and the distribution of the ejecta and the cocoon. Furthermore, it may even hint whether the jet successfully broke out from the ejecta or was choked inside (see e.g. \citealt{Kasliwal2017,Gottlieb2018b}). However, the multiple predictions in the UV/optical/NIR bands rises difficulties in discriminating the potential signals (see also \citealt{Arcavi2018}).

Here we carry out 3D relativistic hydrodynamic (RHD) simulations of a jet propagation in expanding multi-layer ejecta, which we post-process semi-analytically to derive the emissions from sources (iii) and (iv) which have not been examined to date. We also study the effect of the tail ejecta on source (ii) and derive the full light curves for a comparison of all signals. Such an analysis may be of use to infer the origin of a future detected signal.

\section{Numerical Simulations}
\label{sec:numerical}

\subsection{Numerical setup}

We perform two 3D RHD simulations of a jet propagation in expanding ejecta with the code PLUTO v4.0 \citep{Mignone2007}.
The initial setup of the simulations contains a homologous expansion of a two-component ejecta, including a massive ($ M_m = 5\times 10^{-2}\msun $) non-relativistic ($ v<v_m $) component with a radial density profile $ \rho_m(r) r^{-\alpha} $. The massive part is embedded in a faster tail of the ejecta ($ v_m < v < v_t $) which maintains a steeper radial profile $ \rho_t(r)\propto r^{-14} $ \citep{Gottlieb2018b}.
The mass of the latter, $ M_t $, is set by the continuity of the density profile between the two components.
Our choice of the massive ejecta parameters is based on those inferred from GW170817 \citep[see e.g.][]{Kasen2017}.
We assume the free neutrons to reside in the fast tail ejecta and the heavy r-process elements to compose the massive ejecta. To trace each component after it is shocked and mixed with the jet and the cocoon, we employ tracers $ T_m, T_t $ for the massive and tail ejecta, respectively.

We let the ejecta expand for $ t_d $ after the merger, after which we inject a relativistic ($ \Gamma_{0} $) hot ($ \Gamma_{\infty} = 100\Gamma_0 $) top-hat jet at $ z_0 = 3\times 10^8~\rm{cm} $, with an opening angle $ \theta_0 $ and initial radius $ z_0 \theta_0 $. The jet operates for $ t_j $ during which it maintains a constant two-sided luminosity $ L_j $.
We let the simulation run for a fine period $ t_s $, after which most of the mildly-relativistic and relativistic elements reach the homologous expansion phase.
For simulation I we consider typical jet parameters, similar to the ones inferred in GW170817 \citep{Mooley2018}, and adopt favorable tail ejecta properties for the emission. In simulation II we use a wider jet, so that our choice of parameters is slightly more optimal for the jet to break out. The parameters of the simulations are listed in Table \ref{tab_simulations_comparison}.

	\begin{table}
	\setlength{\tabcolsep}{20pt}
	\begin{center}
		
		\begin{tabular}{ | l | c  c | }
			
			\hline
			Simulation & I & II \\ \hline
			$ v_m/c $ & 0.2 & 0.35 \\
			$ \alpha $ & 2 & 3.5 \\
			$ v_t/c $ & $ 0.6 $ & 0.7\\
			$ M_t [\msun] $\footnotemark & $ 4\times 10^{-3} $ & $ 2\times 10^{-5} $\\
			$ \Gamma_0 $ & 5 & 3\\
			$ t_d $[s] & 0.8 & 0.4\\
			$ t_j $[s] & 1 & 2\\
			$ \theta_0 $ & $ 8^\circ $  & $ 15^\circ $\\
			$ L_j [10^{50}\erg\s^{-1}] $ & 1 & 2\\
			$ t_s $[s] & 7.6 & 6.8\\
			
			\hline
			
		\end{tabular}
	\end{center}
\hfill\break

\caption{The parameters in simulations I and II.
		$ v_m $ is the maximal velocity of the massive ejecta component, $ -\alpha $ is the power-law index of its radial density profile, $ v_t $ is the maximal velocity of the tail ejecta, $ M_t $ is the total mass of the tail ejecta, $ \Gamma_0 $ is the initial Lorenz factor of the jet, $ t_d $ is the delay time between the merger and the jet launch, $ t_j $ is the working time of the jet's engine, $ \theta_0 $ is the jet's initial opening angle, $ L_j $ is the jet total luminosity and $ t_s $ is the simulation time.
	}\label{tab_simulations_comparison}
\end{table}
\footnotetext[1]{Simulations of NSMs (e.g. \citealt{Bauswein2013,Hotokezaka2018}) underestimate $ M_m $ compared with the one inferred in GW170817. Normalizing their numerical results to the $ M_m $ in GW170817 produces higher $ M_t $ correspondingly.}
The simulations' grid is Cartesian and a relativistic ideal equation of state with an adiabatic index of 4/3 is applied.
We use three patches along the $ x $ and $ y $ axes, and two patches on the $ z $-axis along which the jet propagates. The inner $ x $ and $ y $ axes are set inside $ |4\times 10^8| $cm with 240 uniform cells. The outer patches are stretched logarithmically to $ |1.5\times 10^{11}| $cm with 280 cells on each side. The $ z $-axis has one uniform patch inside the star from $ z_0 $ to $ 6\times 10^9\cm $ with 400 cells, and anothers logarithmic patch with 1200 cells up to $ 2\times 10^{11} $cm. The total number of cells is thus $ 800\times 800\times 1600 $. This grid resolution is comparable to the one used in \citet{Gottlieb2018a} and was found to converge.

\subsection{Hydrodynamic evolution}

\begin{figure}
	\centering
	\includegraphics[scale=0.23]{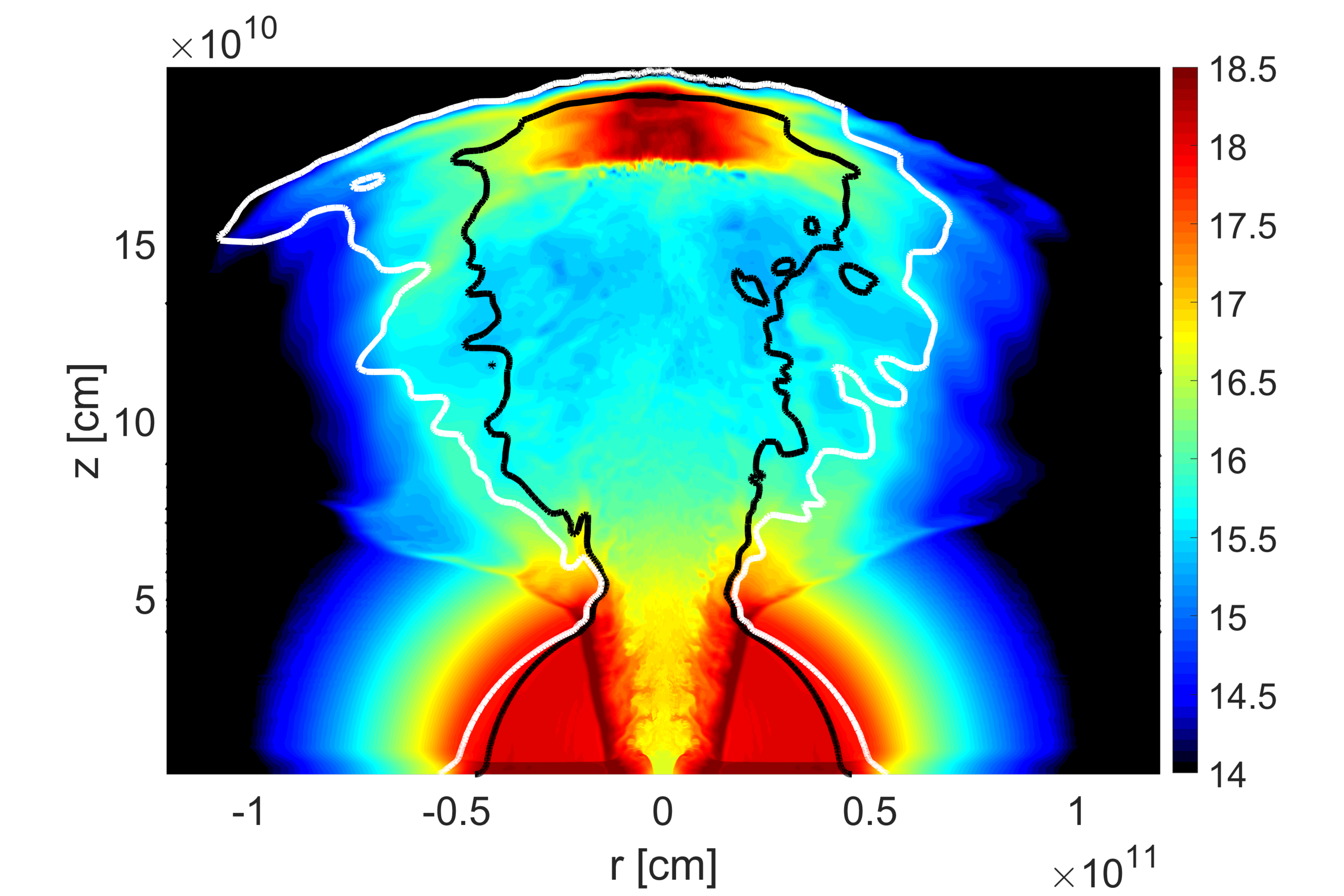}
	\includegraphics[scale=0.23]{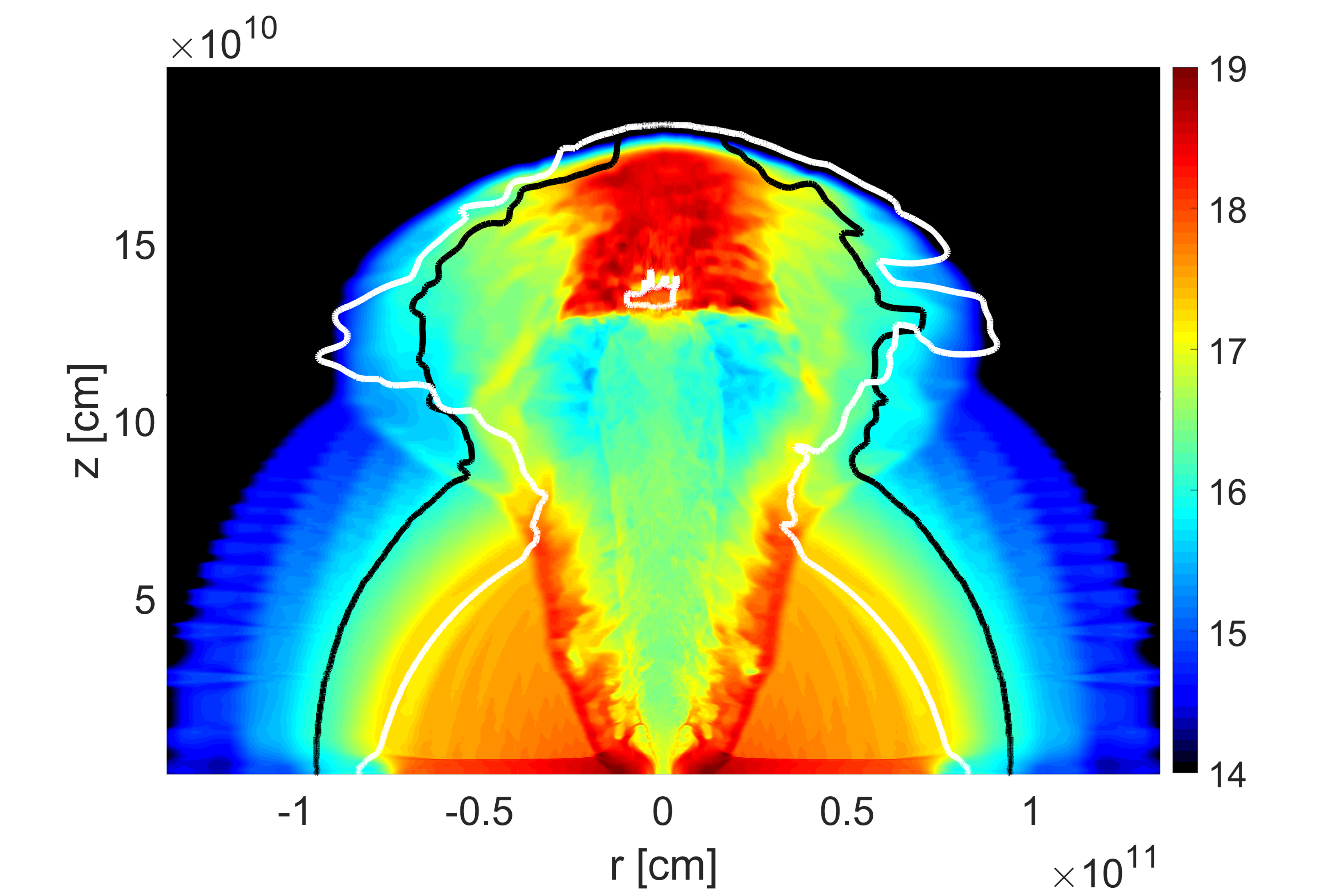}
	\caption[]{
		Logarithmic energy density maps ($ \erg \cm^{-3} $) on the $ x-z $ plane in simulations I (top) and II (bottom), 7.6s and 6.8s after the merger, respectively. The white contours delineate the region within which the fraction of massive ejecta material is above 10\% (the inner white contour in the jet in simulation II represents an unshocked jet region which has not been mixed). The black contour depicts the boundary of the region within which the fast tail ejecta concentration is {\it less} than 10\%.
	}
	\label{fig:map}
\end{figure}

Figure \ref{fig:map} depicts the logarithmic energy density maps at the end of the simulations. 
In both simulations the jet spends $ \sim $1s inside the dense massive ejecta, during which it undergoes intense mixing and forms a hot energetic cocoon ($ E_{c,I} \approx 3\times 10^{49}\erg, E_{c,II} \approx 6\times 10^{49}\erg $).
The interaction between the jet and the massive ejecta along the jet-cocoon interface loads the jet with baryons \citep{Gottlieb2019b} so that both the jet and the cocoon contain a large fraction of heavy r-process elements. This is shown in Figure \ref{fig:map} as the region inside the white contours has at least 10\% of massive ejecta concentration.

As the jet-cocoon structure breaks out from the massive ejecta, it expands freely into the fast tail component.
The tail ejecta is lighter than the massive component and thus its interaction with the jet and the cocoon is weaker. That implies that the jet remains uncontaminated by free neutrons, while the neutrons concentration in the cocoon depends on the specific parameters of the system.
In simulation I, the cocoon is less energetic and the tail ejecta is more massive such that the interaction is stronger \citep{Gottlieb2019b}, resulting in a high fraction of free neutrons in the cocoon ($ > 10\% $ outside the black contours in Figure \ref{fig:map}). Since in simulation II the cocoon is more energetic and the tail ejecta is less massive, the interaction is weaker. As a result, only the less energetic outskirts of the cocoon contain a substantial fraction of free neutrons.

\section{Post-process calculation}
\label{sec:method}

At the end of the simulations most of the outflow expands homologously and we can apply adiabatic relations to infer the hydrodynamics of the system at late times. The extrapolated evolution of the outflow at all times allows us to calculate the UV/optical/NIR light curve during the first day.
We study the contribution of four emission mechanisms: cocoon's cooling, $ \beta $-decay of heavy elements, $ \beta $-decay of free neutrons and synchrotron emission from the $ \beta $-decay electrons.
We consider only emission that originates in the cocoon and the fast tail ejecta, and disregard any emission from the jet or the unshocked massive ejecta component.

For each simulation we consider four models in which we vary the opacity, assumed to be grey opacity $ \kappa $, and the fraction of free neutrons in the tail ejecta, $ \xi $ (Table \ref{tab_models_comparison}).
Note that the fraction of the remaining free neutrons in the tail, $ \xi $, can differ from the initial free neutron fraction, $ 1-2Y_e $, where $ Y_e $ is the electron fraction $ Y_e > 0 $.
For example, it is possible that free neutrons are captured by heavy nuclei during the tail interaction with the cocoon.
However, we note that shocks at high temperature ($ T > 10^{10} $K) alter the fraction of free neutrons by disintegration of heavy elements nuclei to free neutrons and capture of free positrons \citep{Ishii2018}. However, in our simulations the shock driven by the jet maintains $ T \approx 10^{8} $K so that disintegration does not take place and the capture timescale is much longer than the dynamical time \citep{Metzger2015a}. Therefore, the fraction of free neutrons in the cocoon's shocked material is not expected to change following the shock.
		
	\begin{table}
		\setlength{\tabcolsep}{25pt}
		\begin{center}
				
		\begin{tabular}{ | l | c  c | }
			
			\hline
			Model & $ \kappa [\cm^2$/g] & $\xi $ \\ \hline
			$ \A $ & $ 0.1 $ & $ 0.6 $\\
			$ \B $ (Canonical) & $ 1.0 $ & $ 0.6 $ \\
			$ \C $ & $ 10 $  & $ 0.6 $ \\
			$ \D $ & $ 1.0 $ & $ 0.2 $ \\
			
			\hline
			
		\end{tabular}
	\end{center}
	\hfill\break
	
	\caption{The configurations for the emission calculation. $ \kappa $ is the opacity and $ \xi $ is the fraction of free neutrons in the tail ejecta.}\label{tab_models_comparison}
\end{table}

For the calculation of the cocoon's Kilonova and cooling emission we follow \citet{Gottlieb2018a}, where the net radioactive heating rate due to r-process decay is $ e_r\propto t^{-1.3} $ \citep{Metzger2010a}.
For the free neutrons emission which has not been considered in their paper, we apply a similar method. We first calculate the radial optical depth, $ \tau $, along a given solid angle $ (\theta,\phi) $ and at each time $ t $. Then, at each angle and time we find the trapping radius $ r_t $ where $ \tau = c/v $ and the photosphere where $ \tau = 1 $.
The free neutrons heating rate at each time $ t $ and solid angle $ (\theta,\phi) $ is \citep{Kulkarni2005a,Metzger2015a}: 
\begin{equation}\label{key}
e_n(\theta,\phi,t) = \int_{r_t(\theta,\phi,t)}^\infty{3.2\times 10^{14}\frac{\erg}{\s}e^{-t'/\tau_n}m_n(r,\theta,\phi,t)}dr~,
\end{equation}
where $ \tau_n = 881.5\s $ is the free neutrons rest-frame mean lifetime, $ t' = t/\gamma(r,\theta,\phi,t) $ is the comoving time and $ m_n(r,\theta,\phi,t)  = \xi T_t(r,\theta,\phi,t)m(r,\theta,\phi,t) $ is the mass of free neutrons in a given element, $ m $ being the total mass of the element. We assume that the remaining $ (1-\xi)M_t $ in the tail ejecta is not composed of heavy r-process elements (instead, it can either be free protons, alpha particles or other elements which are not radioactively unstable), so that the cocoon's Kilonova is calculated only for $ T_mm $.
At each angle and time the local rest-frame temperature is set at the photosphere and the radiation field is assumed to be a blackbody. Finally, we boost the local emission at all angles and times to the observer frame, where we integrate the total emission.

The free neutrons decay into $ p $, $ \bar{\nu}_e $ and $ e^- $. The decay electrons may power a synchrotron emission if a magnetic field is present. We approximate the energy of the decay electron to be constant by weight-averaging the narrow emitted electron's energy spectrum \citep{Cooper2010}. We find $ E_e \approx 250 $keV, which corresponds to an electron Lorentz factor of $ \gamma_e \approx 1.5 $.
While the strength of the magnetic field in the medium is unknown, it is assumed to reflect some fraction of equipartition with the thermal energy density. First, the ejecta from the NSM is expected to have some initial magnetization (e.g. \citealt{Christie2019,Fernandez2019}). Additionally, the field may also be amplified by different physical processes such as a shock breakout \citep{Waxman2001} of the cocoon from the tail ejecta or the turbulent nature of the hydrodynamic shock (e.g. \citealt{Giacalone2007}). Another possibility for a magnetic field amplification arises if the jet-cocoon structure is magnetized \citep{Medvedev1999}. While our jet-cocoon is unmagnetized, simulations \citep{Gottlieb2020b} show that weakly magnetized jets ($ \sigma = B^2/4\pi\rho c^2 \lesssim 10^{-4} $, $ B $ is the maximal magnetic field of the jet) feature a rather similar cocoon structure with the one obtained in pure hydrodynamic jets. That implies that our analysis is also applicable for such systems, which can naturally induce a magnetic field.

For the synchrotron calculation we calculate in each cell the magnetic field $ B $, neutrons energy $ E_n $, characteristic synchrotron frequency $ \nu_m $, and number of emitted electrons $ N_e $.
Due to the large uncertainty in the value of the magnetic equipartition parameter $ \epsilon_B $, we set $ \epsilon_B = 10^{-6} $ as our conservative canonical value.
For the electron equipartition parameter we use $ \epsilon_e(\theta,\phi,t) = \frac{m_e}{m_p}\frac{\gamma_e-1}{\gamma_n(\theta,\phi,t)-1} $, where $ m_e $ and $ m_p $ are the electron and proton masses, respectively, and $ \gamma_n $ is the Lorentz factor of the neutrons.
Then, for each element we calculate the maximal spectral luminosity, $ L_{\nu,{\rm max}} = N_eB\sigma_Tm_ec^2/3q_e $, where $ \sigma_T $ and $ q_e $ are the Thomson cross section and electron charge, respectively.
At the relevant timescales self-absorption is expected to take place as $ \nu_a > \nu_m $, where the self-absorption frequency at time $ t $ and solid angle $ (\theta,\phi) $ is \citep{Waxman1997}:
\begin{equation}\label{eq:nu_a}
\nu_a(\theta,\phi,t) = \frac{1.2}{\epsilon_e(\theta,\phi,t)}\Big(E_{n,{\rm iso,49}}(\theta,\phi,t)\epsilon_B n_0(\theta,\phi,t)^3\Big)^{0.2}\rm{GHz}~,
\end{equation}
where $ n $ is the medium number density at $ r_t $, $ Q_x $ denotes the value of the quantity $ Q $ in units of $ 10^x $ times its c.g.s. units.
We find the spectral regime of each element with respect to $ \nu_m, \nu_a $ and the cooling frequency $ \nu_c $ to calculate the local spectral luminosity $ L_\nu $ in the comoving frame which is then boosted and integrated in the observer frame.

\section{Light curves}
\label{sec:results}

\begin{figure*}
	\centering
	\includegraphics[scale=0.35,trim=1cm 0cm 0cm 0cm]{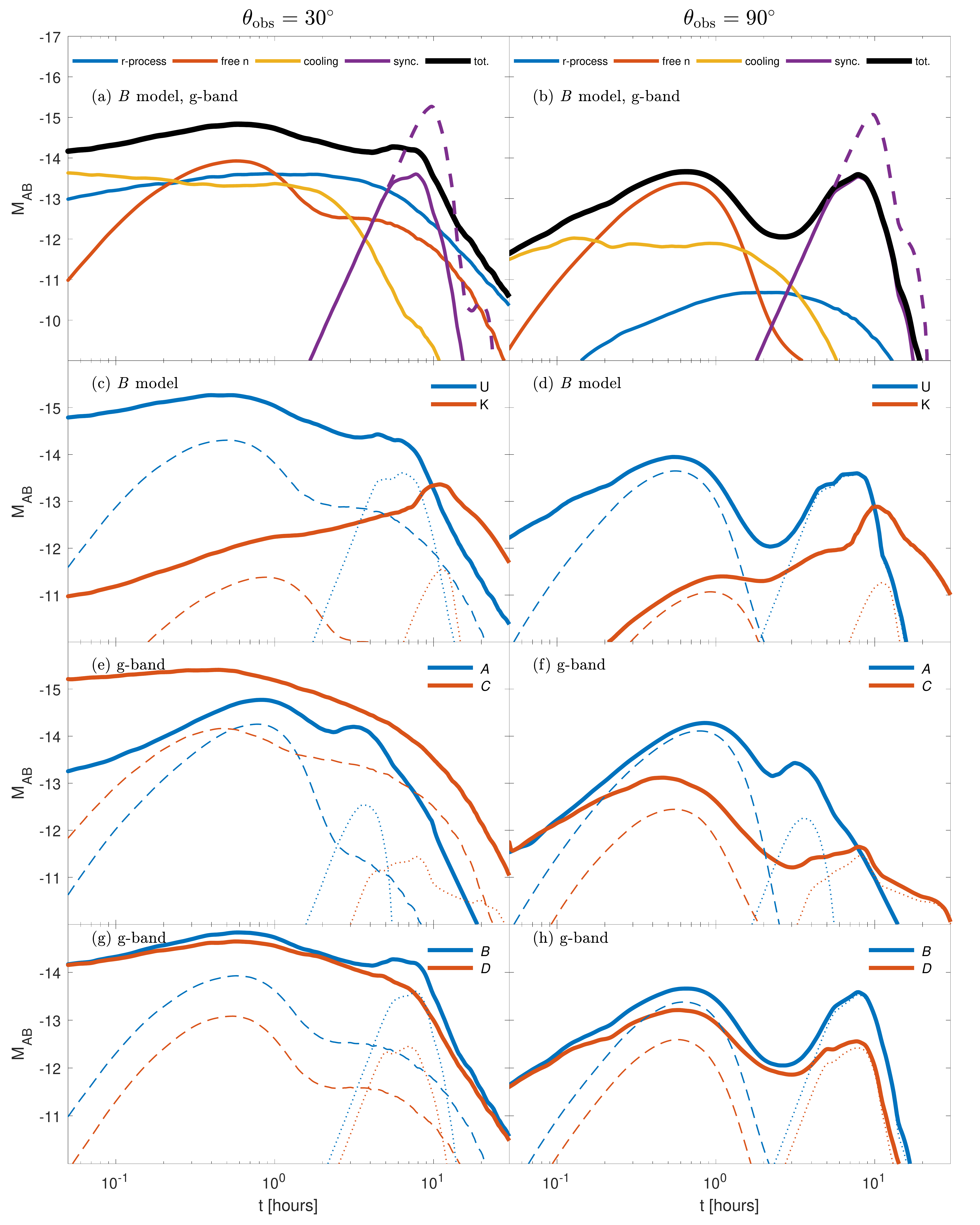}
	\caption[]{
		UV/optical/NIR light curves from simulation I at $ \thobs = 30^\circ $ (left) and $ \thobs = 90^\circ $ (right).
		Panels (a,b) show the contribution of each emission mechanism in the canonical model, $ \B $, at the g-band. The dashed purple line depicts the synchrotron emission with $ \epsilon_B = 10^{-4} $.
		Panels (c,d) depict the total emission in the U and K bands.
		Panels (e,f) show the total g-band emission for lower and higher opacities in models $ \A $ and $ \C $, respectively.
		Panels (g,h) demonstrate the effect of the free neutrons fraction, $ \xi $, on the emission with models $ \B $ and $ \D $.
		In panels (c-h) dashed and dotted lines reflect the free neutron decay and resulting electron synchrotron emissions, respectively.
	}
	\label{fig:optical1}
\end{figure*}

\begin{figure*}
	\centering
	\includegraphics[scale=0.35,trim=1cm 0cm 0cm 0cm]{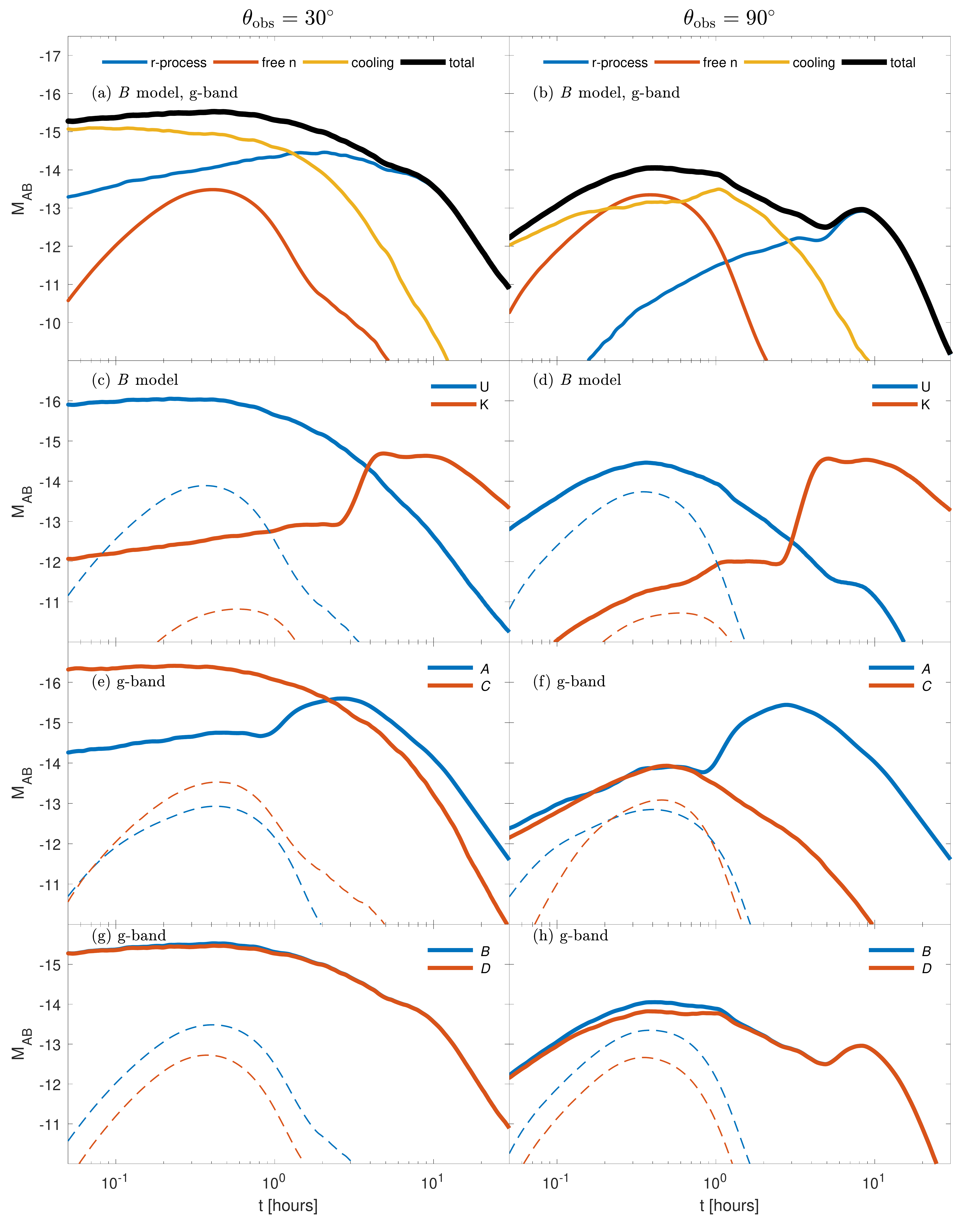}
	\caption[]{
		Same as Figure \ref{fig:optical1} for simulation II. Panels (g,h) show that unlike simulation I, the total light curves are insensitive to the value of $ \xi $ as the free neutrons decay and synchrotron are negligible compared to cooling and r-process decay emissions, owing to the small tail mass.
	}
	\label{fig:optical2}
\end{figure*}

Figures \ref{fig:optical1} and \ref{fig:optical2} depict the numerical UV/optical/NIR light curves for simulations I and II, respectively. These are shown by emission mechanism contribution (panels a,b), at a variety of bands (c,d), and for different opacities (e,f) and free neutrons fractions (g,h).
The presence of the cocoon at high latitudes alters the spatial distribution of the ejecta and introduces relativistic effects, thereby leading to differences between small and large viewing angles. We find that the light curves can be roughly divided into two regions: for observers who are within $ \sim 1/\Gamma\beta \approx 25^\circ $ from the cocoon's opening angle $ \theta \approx 30^\circ $ and outside of it. At small viewing angles $ \thobs \lesssim 50^\circ $ the signal is similarly enhanced due to the cocoon's mildly-relativistic motion (left panels). For all $ \thobs \gtrsim 60^\circ $ the cocoon's contribution is secondary (right panels).

\subsection{First hour emission}

At early times ($ \lesssim 1$ hour) the temperature is high ($ T\approx 6\times 10^4 $K after twenty minutes) and the emission lies in the Rayleigh-Jeans tail such that the UV signal is the brightest. The relevant emission sources at these times are free neutrons decay and cocoon's cooling emission.
We find that without relativistic effects, simulation I, in which the tail ejecta is more massive by a factor of 200, would have featured a stronger free neutrons emission (the difference is however small, by a mere one absolute magnitude, see \citealt{Nakar2019}). In simulation II the tail ejecta is faster so that its boost places the free neutrons emission to be similar in both simulations with a UV peak at $ \rm{M_{AB}} \approx -14 $ (also similar to the peak obtained by \citealt{Metzger2015a}). That implies that for the neutrons mass in simulation I and their velocity in simulation II, the UV signal will peak at an absolute magnitude $ \rm{M_{AB}} \approx -15 $.
At small viewing angles the free neutrons emission is boosted by the cocoon and magnified by up to one absolute magnitude in simulation I. In simulation II the fast tail is moving faster and thus the boost by the cocoon is less significant.

The main difference in the cooling emission between the simulations lies in the cocoon's energy. As $ E_{c,2} \approx 2 E_{c,1} $, the cooling emission in simulation II is $ \sim 1 $ magnitude brighter. In both simulations we find that a large part of the cocoon is beamed away from observers at large $ \thobs $, resulting in $ \sim 2 $ magnitudes fainter signal compared to small $ \thobs $.
The shape of the cooling emission is dictated by the luminosity and temperature power-laws. We find that the temperature drops as $ T \propto t^{-0.7} $, and the cooling luminosity drops fast as $ L \propto t^{-2} $ at $ \thobs=30^\circ $ and $ L\propto t^{-1.3} $ at $ \thobs=90^\circ $. From the Rayleigh-Jeans regime, $ L_\nu \propto LT^{-3} $, it follows that the cooling spectral luminosity $ L_\nu $ is roughly constant as long as the emission is in this regime.

Panels (e,f) show that the opacity of the matter also affects both the absolute and the relative contribution of each emission mechanism. Low opacity leads to an early rise of cooling emission when the temperature is still high and thus the UV/optical luminosity is low, whereas high opacity of $ \kappa \approx 10\cm^2 $/g yields the strongest cooling emission, up to a UV peak at $ \rm{M_{AB}} \approx -17 $.
The sensitivity of the free neutrons to the opacity depends on the mass and the viewing angle.
In simulation I the fast tail is sufficiently massive such that it may be optically thick at low latitudes. Consequently, lower opacities for which there are less adiabatic loses, produce the strongest signals.
At small viewing angles and light tail ejecta (simulation II) the free neutrons component becomes optically thin early on so that the opacity does not affect the light curves. 
We conclude that both the boost and the opacity play a more important role for the cocoon's cooling than the free neutrons emission.
As simulations I and II show, the dominating emission mechanism during the first hour depends on the fast tail characteristics, cocoon's energy, opacity and viewing angle.

\subsection{Few hours emission}

The shocked ejecta in the cocoon generates a cocoon's Kilonova signal that peaks in the NIR a few hours after the merger. This emission may be attenuated if the fast tail component contains a negligible fraction of r-process elements and is sufficiently massive to obscure the slower parts of the heavy shocked ejecta.
In simulation II the cocoon's Kilonova is bright ($ {\rm M_{AB}} \sim -15 $) and quasi-isotropic owing to the optically thinness of the fast component, similar to what \citet{Gottlieb2018a} found. In simulation I however the fast tail component becomes optically thin only at
\begin{equation}\label{eq:tau}
t \approx 3 \Bigg(\frac{\kappa_0 M_{t,29}}{\beta_{t,-0.5}^2}\Bigg)^{0.5} {\rm hours} ~,
\end{equation}
where $ \beta_t \equiv v_t/c $. Therefore only the mildly-relativistic component at high latitudes contributes to the cocoon's Kilonova.
It then follows that in simulation I the emission is fainter and non-isotropic where at large $ \thobs $ the emission is fainter by 3 magnitudes compared to low $ \thobs $.
The shape of the emission is dictated by the luminosity and temperature dependencies. We find that at all relevant times and for most observers and models, the total bolometric luminosity drops fast as $ L \approx 10^{42}(t/{\rm 1hr})^{-1.6} {\erg \s^{-1}} $, leading to a quasi-flat spectral luminosity at early times, similar to the cooling emission.
As expected, the difference between models $ \A $ and $ \C $ points at the importance of the opacity for the cocoon's Kilonova brightness. Equation (\ref{eq:tau}) shows that outflows with a lower opacity reach their photosphere earlier and suffer from less adiabatic loses, thereby yielding a peak emission that is $ \gtrsim 3 $ absolute magnitudes brighter than ones with high $ \kappa $ (see also \citealt{Gottlieb2018a}).

If the free neutrons mass is high enough, a synchrotron emission on hours timescale emerges, owing to self-absorption processes as $ \nu_m < \nu < \nu_a $.
The peak of the signal is obtained when either $ \nu_a $ drops below $ \nu $ or all free neutrons reach their photosphere, whichever comes first.
Equation (\ref{eq:tau}) dictates that for the canonical model in simulations I and II all the free neutrons reach the photosphere after $ \sim $ 30 hours and $ \sim 2 $ hours, respectively.
During the adiabatic expansion $ n \propto t^{-3} $ and before all the free neutrons reach the photosphere, the total mass of the revealed neutrons increases somewhat slower than $ t^2\kappa^{-1} $ \citep{Piran2013}. It then follows from Equation (\ref{eq:nu_a}) that the self-absorption frequency evolves as
\begin{equation}\label{eq:nu_at}
\nu_a \approx \nu_{a,0}\bigg(\frac{\epsilon_{B,-6}}{\kappa_0}\bigg)^{0.2}t_0^{-1.5}~,
\end{equation}
where from Equation (\ref{eq:nu_a}) $ \nu_{a,0} \approx 10^{22} $Hz is the self-absorption frequency 1s after the merger. We therefore get that $ \nu_a = \nu $ at,
\begin{equation}\label{eq:nu_ta}
t = \bigg[\frac{\nu_{a,0}}{\nu}\Big(\frac{\epsilon_{B,-6}}{\kappa_0}\Big)^{0.2}\bigg]^{2/3}\s\approx 10~{\rm hours}~,
\end{equation}
assuming the parameters of our canonical model and U band frequency.
Finally, it follows from Equation (\ref{eq:nu_at}) that the spectral luminosity at $ \nu_m < \nu < \nu_a $ rises as,
\begin{equation}\label{eq:Lv}
L_\nu \propto N_e\epsilon_B\Big(\frac{\nu}{\nu_a}\Big)^{2.5} \propto N_e (\epsilon_B\kappa)^{0.5}\nu^{2.5}t^{3.75}~.
\end{equation}

The fast rise of $ L_\nu $, as seen in Equation (\ref{eq:Lv}) and the light curves of simulation I, implies that the peak emission strongly depends on the time of the peak. Equation (\ref{eq:nu_ta}) shows that the UV/optical peak is obtained about 10 hours after the merger, as seen in the light curves of simulation I. This signal is quasi-isotropic and thus the synchrotron emission is more prominent at large viewing angles where the cocoon's Kilonova is fainter.
Equation (\ref{eq:Lv}) also shows that higher values of $ \epsilon_B $ (panels a,b), $ \nu $ (c,d; up to $ \nu_c $), $ \kappa $ (e,f) or $ \xi $ (g,h) yield a brighter synchrotron emission. Note however that Equation (\ref{eq:tau}) indicates that higher opacities result in late exposure of all free neutrons, so that $ N_e $ is lower. For example, in simulation I, $ \kappa = 1\cm^2 $/g produces the brightest emission as for higher opacities a smaller fraction of neutrons reaches the photosphere before $ \nu > \nu_a $.
For simulation II all neutrons reach their photosphere after $ \sim $ hour, much before $ \nu_a = \nu $. At this time the emission is too faint to be detected.
Table \ref{tab:synchrotron} summarizes the dependencies of the synchrotron timescales in Equations (\ref{eq:tau}) and (\ref{eq:nu_ta}) and peak magnitudes (Eq. (\ref{eq:Lv})) on the free neutrons fraction, tail mass and opacity used in our different models.

Generalizing the synchrotron emission, we find that for typical values of $ \kappa $ and $ v_t $, the condition for a detectable signal is set by Equation (\ref{eq:tau}). It then follows that a UV/optical signal emerges at $ M_{\rm AB} \gtrsim -12 $ if the free neutrons mass is
$ M_n \gtrsim 10^{-4}\msun\xi_{-0.2}^{0.6}\kappa_0^{-0.8}\beta_{-0.3}^{1.3} $, assuming all decay electrons have been revealed by the time of the peak. Differentiating between the late signals of the cocoon's Kilonova and the synchrotron emission can be achieved by multi-wavelength observations as the latter rises in higher $ \nu $ and vice versa.

	\begin{table}
	\setlength{\tabcolsep}{8pt}
		\begin{center}

		\begin{tabular}{ | l | c  c  c | }

			\hline
			Sim./Model & Eq. \ref{eq:tau} [hours] & Eq. \ref{eq:nu_ta} [hours] & Eq. \ref{eq:Lv} [$ M_{\rm {AB}} $] \\ \hline
			 I/a   & 4   & 14 & -12\\
			 I/b   & 8   & 10 & -14\\
			 I/c   & 27  & 6  & -11\\
			 I/d   & 8   & 10 & -12\\
			 II/a  & 0.3 & 14 & -1\\
			 II/b  & 1   & 10 & -3\\
			 II/c  & 3   & 7  & -5\\
			 II/d  & 1   & 10 & -1\\
			\hline

		\end{tabular}
	\end{center}
	\hfill\break
	
	\caption{Timescales and peak absolute magnitudes of the different models, obtained by Equations (\ref{eq:tau}), (\ref{eq:nu_ta}) and (\ref{eq:Lv}). The absolute magnitude in the third column is calculated at the time of the peak, that is the minimal time between the first two columns.
		}\label{tab:synchrotron}
\end{table}

\section{Conclusions}
\label{sec:conclusions}

Simulations of a double NSM have shown that a significant amount of free neutrons is ejected following the merger \citep{Bauswein2013}. The $ \beta $-decay of the free neutrons after $ \sim 15 $ minutes may power an early bright UV/optical signal \citep{Kulkarni2005a,Metzger2015a}.
In this paper, we studied the effect of the cocoon on the free neutrons emission and the synchrotron emission from the $ \beta $-decay electrons. We performed 3D RHD simulations of a jet-cocoon propagation in a multi-layer expanding ejecta from the merger. We post-processed the numerical results to calculate these signals and compare them with the expected cocoon's Kilonova and cooling emission (\citealt{Gottlieb2018a}).
Inferring the signatures of these signals in future observations can be used to constrain the physical characteristics of the ejecta and the cocoon.

During the first hour the dominating sources of emission are cocoon's cooling and free neutrons decay.
Our calculations show that the cocoon can boost the free neutrons emission by up to $ \sim 1 $ absolute magnitude to power a bright UV peak at $ {\rm M_{AB}} \approx $ -15 to -16 at small viewing angles.
Two main differences are found between the two emission sources:
(i) While for a quasi-isotropic ejecta of free neutrons a prominent UV signal ($ {\rm M_{AB}} \approx $ -14) rises even at large viewing angles, the cooling emission strongly depends on the viewing angle; and
(ii) The cooling emission is also greatly affected by the opacity, growing brighter for higher opacities. The free neutrons emission is sensitive to the opacity only at low latitudes, where there is enough neutrons mass to be optically thick in the relevant timescales. In such cases the emission is brighter at lower opacities, owing to the lower adiabatic loses.
As both signals can share comparable peak times and fluxes, the dominant emission source depends on the specific parameters of the system.

A few hours after the merger the cocoon's Kilonova and synchrotron emission dominate the light curves.
We find that for high, but not unreasonable mass of free neutrons,
$ M_n \gtrsim 10^{-4}\msun\xi_{-0.2}^{0.6}\kappa_0^{-0.8}\beta_{-0.3}^{1.3} $, 
a quasi-isotropic synchrotron UV/optical signal emerges $ \sim 8 $ hours after the merger at $ M_{\rm AB} \gtrsim $ -12, owing to synchrotron self-absorption. The peak emission depends on the highly uncertain value of the magnetic equipartition parameter $ \epsilon_B $. For a reasonable $ \epsilon_B = 10^{-4} $, the peak emission of our canonical model in simulation I is obtained at $ {\rm M_{AB}} \approx $ -15.
The cocoon's Kilonova emerges in the NIR bands if the opacity is not too high, $ \kappa \lesssim 1\cm^2 $/g.
However, if the tail ejecta is massive, the optical thickness of the unshocked tail ejecta at low latitudes will partly obscure the cocoon's Kilonova. If the fast tail is optically thin at these times, the cocoon's Kilonova signal is also quasi-isotropic.
A multi-wavelength observations on a few hours timescale will enable to separate the synchrotron emission from the cocoon's Kilonova signal.
Considering the bright synchrotron emission, even for small $ \epsilon_B $, a null detection of a UV signal on hours timescale may place constraints on the total ejected mass in free neutrons.

Our results show that not only that the free neutrons properties alter the free neutrons and synchrotron signals, but affect all emission sources.
Unfortunately, the mass and velocity of the free neutrons are highly uncertain.
Recently, \citet{Ishii2018} found that the mass of shock-heated free neutrons released from the outer crust of the NS is substantially lower, $ M_n \approx 10^{-6}\msun $, than what has been found in previous simulations.
\citet{Ishii2018} suggested that the discrepancy can be resolved if the rest of the free neutrons originate from tidal debris or $ \nu $-driven wind. For the former it implies that most of the free neutrons lie at low latitudes and have a minimal interaction with the cocoon.
As mentioned above, a lighter mass of free neutrons will naturally reduce the emission from free neutrons but also allows a bright cocoon's Kilonova emission by virtue of it being optically thin.
The velocity of the mildly-relativistic component also plays a crucial role as the Doppler factor grows larger and relativistic effects can substantially enhance the emission.

Finally, while our simulations contain jets that successfully break out from the expanding ejecta, some jets may never break out, leaving the cocoon as the only relativistic component. In such scenarios the cocoon opens up to wider opening angles, thereby becoming more massive and slower. Related calculations of the cocoon's cooling emission and Kilonova \citep{Kasliwal2017,Nakar2018} have shown that such systems yield a longer and brighter signals, owing to the cocoon's characteristics.
Since a choked jet's cocoon typically moves at a velocity that is comparable to the tail ejecta, we expect that unlike other sources of signals, the effect of such a cocoon on the free neutrons emission will be only mild.

\section*{Acknowledgements}

We thank Ehud Nakar and the anonymous referee for helpful comments.
This work was supported in part by Harvard's Black Hole Initiative which is funded by grants from JTF and GBMF.

\bibliographystyle{mnras}
\bibliography{../Cocoon-papers/Free_neutrons}

\appendix

\label{lastpage}
\end{document}